\documentstyle{article}
\title{\sc Complementary Relativity}
\author{ Pedro F. Gonz\'alez-D\'{\i}az.\\
Instituto de Matem\'aticas y F\'{\i}sica Fundamental\\
Consejo Superior de Investigaciones Cientificas\\
Serrano 121, 28006 Madrid (SPAIN)\\
}
\date{April 7, 1994}
\begin{document}
\maketitle
\large
\setlength{\baselineskip}{0.5cm}

There are two possible arenas where events may occur - that
of customary space-time and that of a vacuum where events
cannot be labelled by space and time, but by a continuum
of momentum-energy coordinates. Special theory of
relativity has been formulated in a vacuum momentum-energy
representation which is equivalent to Einstein special
theory of relativity and predicts just the same results
as it. Although in this sense such a formulation would
be at least classically useless, its consistent
extension to noninertial frames produces a
momentum-energy metric which behaves as a new
dynamical quantity that is here interpreted in terms
of a cosmological field. This new field would give
rise to interactions transmitted through vacuum,
rather than space-time, and is complementary to
gravity in that its strength would vary inversely to
as that of gravity does. Using a strong-field
approximation, we suggest that the existence of these
cosmological interactions would induce a shift of
luminous energy which could justify the existence
of all the assumed invisible matter in the universe,
so as the high luminosities found in active galactic
nuclei and quasars.

\pagebreak

The general framework of theoretical physics may be incomplete in
one respect. Whereas quantum mechanics possesses two fully
equivalent, though complementary, representations,
only the representation that would correspond
to position has been worked out in relativity.
Although a formulation of special relativity
in momentum representation
would be fully equivalent to Einstein special
relativity in that it predicted exactly the
same results,
the lack of that representation in general
relativity might be hiding some new physics.

In Einstein special relativity, the Minkowskian
coordinates $t$ and $x$ have a double function:
they serve as labels, but at the same time they
also inform us, through Lorentz transformations, about
actual time durations and space distances,
measurable with clocks and meters. Moreover,
although in Einstein relativity momentum and
energy can never label real events, they are
obtained as actual quantities in its associated
relativistic mechanics. Therefore, if we attempted
to build up a consistent momentum-energy
representation for special relativity, one would
expect {\it no} new predictions from it, but just
again the same expressions for time duration and
space distances (given in this case as mechanical
quantities), now with the coordinates $p$ and $e$
having the double function of serving as labels
and informing us about the same actual values of momentum
and energy that are predicted by Einstein kinematics.
It is in this sense that such a momentum-energy
representation of special relativity
would be useless.

However, in general relativity this situation could
change drastically. For a given representation, we
must distinguish here between coordinate labels from
proper intervals, entering at the two totally different
levels that correspond, respectively, to differential
topology and metric geometry. In the usual space-time
representation of general relativity, showings of a
physical clock are predicted not only by the labels
that distinguish events, but also by the metric, and
the metric describes at the same time a dynamical
quantity - the gravitational field. Likewise, in a
consistent momentum-energy representation of general
relativity, besides mechanical time durations and
space distances, one would expect the emergence of
an essentially new quantity - the metric of the
momentum-energy continuum, which would help, together
the coordinate labels, to construct actual momentum
and energy intervals. Such a momentum-energy metric
would describe at the same time a new "dynamical"
quantity: a field which should be complementary to
gravity. It is the physics associated with this new
field what may be gained from a complete formulation
of relativity.

In what follows I will formulate a momentum-energy
representation for relativity. In order for the
resulting full theory to be self-consistent, such a
formulation should satisfy the following requirements.

(i) The kinematics of special relativity
(i.e. the relations between coordinate
labels) in the momentum-energy representation must
satisfy all mechanical Einstein four-momentum transformations,
and its associated mechanics (i.e. the quantities derived
from an action principle)
must in turn obey the usual
Lorentz transformations.

(ii) Since vacuum is not but what remains after eliminating
space-time, we must take it as the arena where events can
be continuously labelled by momentum and energy coordinates.
Thus, whereas description of a given system in space-time
implies that such a system is just a space-time
part from a necessarily larger system where at least
an external observer is also included, its description
in the vacuum momentum-energy continuum requires
considering the system as a space-temporal whole where
even observers are included, so that no independent
evolution of any space-time part of it is possible.

(iii) The nonrelativistic limit $c\rightarrow 0$ of the
resulting mechanical relations between time durations
and space distances should either produce trivial or
known results, or not exist at all. The nonrelativistic
limit of the kinematical transformations of momentum
and energy must predict values of the vacuum energy
which depend on the chosen reference system, and
values of the vacuum momentum such that this behaved
as an absolute quantity.

The latter requirement needs some further explanation.
Consider a system {\it S} which evolves uniformly
(i.e. at a constant rate $\frac{dp}{de}$) in the vacuum
momentum-energy continuum. Since, after requirement (i),
its evolution rate is $\frac{dp}{de}=\frac{v}{c^{2}}$,
we can see why the components of momentum must become
absolute quantities in the nonrelativistic limit, where
energy will still depend on the bare velocity $v$. For
in such a limit, one would not expect the system {\it S}
with energy $e_{1}$ to interact with itseft with a
different energy $e_{2}$ because, then, the maximum rate
of signal propagation in momentum-energy, $\frac{1}{c}$,
becomes zero. This is precisely what one should expect
from the zero-point nonrelativistic oscillator contribution:
a nonzero, arbitrary energy and a strictly zero momentum.

Passing to the domain where $c$ is finite, we see that
the maximum rate of signal propagation in momentum-energy
is no longer zero and, therefore, the momentum components
become no longer absolute quantities. This will give rise
to the emergence of a purely relativistic interaction
of the system {\it S} with itseft when its energy changes
along evolution.
We can then introduce momentum-energy reference systems evolving
uniformly relative to each other with relative rates $\frac{v}{c^{2}}$,
so as an extended principle of relativity according to which all the laws
of nature are identical in all "inertial" momentum-energy reference
systems, if the equations expressing the laws and the events that
take place in such reference systems are all described in terms of
momenta and energies. Such laws must then be invariant with respect
to transformations of momenta and energies from one momentum-energy
reference system to another.

A differential interval defined in one of such reference systems
can be given by
\begin{equation}
dn^{2}=\frac{de^{2}}{c^{2}}-dp_{x}^{2}-dp_{y}^{2}-dp_{z}^{2}.
\end{equation}
The principle of relativity for momentum-energy continuum implies
that $dn$ will be the same in all inertial momentum-energy systems,
and leads to the definition of a proper energy given by
\begin{equation}
de=\frac{de'}{\gamma},  \gamma = (1-\frac{v^{2}}{c^{2}})^{\frac{1}{2}}.
\end{equation}

Let us consider two inertial momentum-energy reference
systems independently evolving with a relative rate
$\frac{v}{c^{2}}$. From the above discussion it follows
that if the energy origin is chosen at the point where
both systems coincide, and such systems evolve so that
their $p_{x}$-axes always coincide, then we will have
in the limit $c\rightarrow 0$
\begin{equation}
p_{x}=p_{x}', p_{y}=p_{y}', p_{z}=p_{z}', e=e'+p_{x}v .
\end{equation}
On the other hand, if $c$ is kept finite,
it is easy to see that the transformations that leave invariant the
interval are
\begin{equation}
p_{x}=\frac{p_{x}'+\frac{v}{c^{2}}e'}{\gamma}, p_{y}=p_{y}', p_{z}=p_{z}',
e=\frac{p_{x}'v+e'}{\gamma},
\end{equation}
which, in turn, coincide with the transformation formulas for momentum energy
4-vector of Einstein relativistic mechanics. Equations (4) lead to expressions
for the transformations of velocities, general 4-vectors, and unit
4-velocities, which exactly coincide with those of Einstein relativistic
kinematics, and reduce to (3) as $c\rightarrow 0$. Thus, the
transformations (4) do satisfy the kinematical parts of the
requirements in (i) and (iii).

In order to formulate the relativistic mechanics in
momentum-energy representation,
let us consider a free system
evolving in the momentum-energy continuum. For such a system there
should exist a certain integral (the counterpart to action of
Einstein relativity) which has the minimum value for actual evolution
of the system in the momentum-energy continuum. This integral must
have the form
\begin{equation}
P=-\beta\int_{a}^{b}dn=-\frac{\beta}{c}\int_{e_{1}}^{e_{2}}de\gamma
=\int_{e_{1}}^{e_{2}}\tilde{L}de,
\end{equation}
where $\int_{a}^{b}$ is an integral along a momentum-energy world line
of the system between two particular events characterizing the
momentum of the system when it has energies $e_{1}$ and $e_{2}$,
and $\beta$ is some constant that characterizes the system. The
coefficient $\tilde{L}$ of $de$ plays the role of a Lagrangian and
has the physical dimensions of a time. For $P$ to have the
dimensions of an action, unlike Einstein relativity
where each system is characterized by its rest energy $mc^{2}$, here
each system should be characterized by the complementary quantity
to its rest energy, that is its rest time $T_{0}$. We take therefore
$\beta=cT_{0}$, and hence the integral $P$ for a free
temporal system becomes
\begin{equation}
P=-T_{0}\int_{e_{1}}^{e_{2}}de\gamma ,
\end{equation}
with $\tilde{L}=-T_{0}\gamma$.

Instead of a momentum and an energy, the mechanical system will now
be described by a space distance $R$ and a time duration $T$.
Assuming the momentum-energy coordinate space to be homogeneous,
so that the properties of the system remain invariant under
infinitesimal parallel displacements of rate $\frac{v}{c^{2}}$
and energy $e$, the quantities $R$ and $T$ would be conseved
and can be obtained using the same Lagrangian
principles as in classical mechanics, but in our complementary
representation, i.e.
\begin{equation}
R=\frac{\partial\tilde{L}}{\partial(\frac{v}{c^{2}})}=\frac{T_{0}v}{\gamma},
T=R\frac{v}{c^{2}}-\tilde{L}=\frac{T_{0}}{\gamma}.
\end{equation}

We have to check that the relativistic mechanics expressed by (5)-(7)
is consistent with the full relativistic picture, i.e. we have to
check that by substituting space distance
and time duration given in (7), expressed as a
4-vector in terms of the corresponding 4-velocity, in the
transformation formulas for a general 4-vector, one obtains usual
Lorentz transformations. That this is indeed the case can be readily
seen by using the principle of least action, $\delta P=0$, and
$dn=(dp_{i}dp^{i})^{\frac{1}{2}}$, with $p^{0}=\frac{e}{c}$, $p^{1}
=p_{x}$, $p^{2}=p_{y}$, $p^{3}=p_{z}$. We then obtain $\delta P=
-T_{0}s_{i}\delta p^{i}$, where $s_{i}=\frac{dp^{i}}{dn}=u_{i}$,
$u_{i}$ being the Einstein unit 4-velocity. It follows that
\begin{equation}
x_{i}=-\frac{\partial P}{\partial p^{i}}=(cT,R)=T_{0}u_{i}
\end{equation}
is the distance 4-vector. It turns out that the square of the
length of momentum 4-vector, $(p^{0})^{2}-\sum_{\alpha=1}^{3}(p^{\alpha})^{2}$,
is invariant under transformations (4). Generalizing to any 4-vector
$A^{i}$ which transforms like the components of the momentum
4-vector under (4), we recover the usual transformation formulas
for 4-vectors of Einstein relativity. It is now inmediately seen
that by substituting (8) into such formulas one obtains usual
Lorentz transformations. This completes fullfilment of
requirement (i).

We also note that the formula for $T$ in
(7) has no general nonrelativistic counterpart. In fact, in the
limit $c\rightarrow\infty$, we obtain from (7)
\begin{equation}
T\approx T_{0}+\frac{T_{0}v^{2}}{2c^{2}}\approx T_{0}, R\approx T_{0}v,
\end{equation}
i.e. the nonrelativistic limit of $T$ and $R$ reduces, respectively,
to the rest time and a distance-velocity law which may be
trivially interpreted as the customary definition of velocity.
Rest time $T_{0}$ would correspond to the period of the
wave that quantum theory associates with the corpuscular
characteristics of the system.

On the other hand, it follows from (7)
\begin{equation}
R=Tv
\end{equation}
\begin{equation}
T^{2}c^{2}=R^{2}+T_{0}^{2}c^{2}.
\end{equation}
Expression (9) should now correspond to the relativistic expression for
the definition of velocity.
We finally note that (11) must correspond to the analogue of
the usual relativistic Hamiltonian in our complementary formalism
for relativity. If we express time $T$ in terms of the distances $R$, then
we have a complementary relativistic Hamiltonian
\begin{equation}
T\equiv H_{T}=\frac{1}{c}(R^{2}+T_{0}^{2}c^{2})^{\frac{1}{2}},
\end{equation}
which has the physical dimension of a time. Law (12) must correspond
to the Minkowskian function $F$ which is the conjugate counterpart
to Hamiltonian and whose existence has been recently suggested [1].

We still have to check that our mechanical relation (11)
satisfies requirement (iii).
In the limit $c\rightarrow\infty$ (11) again gives rise to no
nonrelativistic complementary Hamiltonian, except for the rest
time of the system. We conclude therefore that law (12) cannot
have any mechanical effects in our familiar low-velocity physics,
other than making it possible the existence of the
quantum-mechanical wave-particle dual character for
the given system.
When applied to the whole universe, this law will describe the
cosmological evolution in the vacuum momentum-energy continuum.
A quantum-mechanical wave equation can also be derived from (11)
by introducing the operators $\hat{T}=i\hbar\frac{\delta}{\delta e}$,
$\hat{R}=i\hbar\frac{\delta}{\delta p}$. Using a wave function
$\Upsilon\equiv\Upsilon(p,e)$, we obtain
\begin{equation}
-\hbar^{2}\frac{\partial^{2}\Upsilon}{\partial
e^{2}}=\frac{1}{c^{2}}(-\hbar^{2}\frac{\partial^{2}}{\partial
p^{2}}+T_{0}^{2}c^{2}+V(p))\Upsilon,
\end{equation}
where we have introduced a generic potential $V(p)$. This is the
complementary-relativity analogue to the Klein-Gordon equation. If,
as it is the case for the whole universe, the system is closed, then
one would expect a discrete $T$-spectrum which would associate with
an infinite set of universes frozen at the given values of $T$. This
spectrum would only become continuous in the classical region that
corresponded to very large values of $T$. We finally note that
the quantum description of systems that show
time asymmetry could only be accounted for whenever we assume a
haft-integer intrinsic angular momentum for the whole system, so that,
instead of (13), one would have a Dirac-like wave equation
\begin{equation}
(\gamma^{i}\frac{\partial}{\partial p^{i}}+cT_{0}+V(p))\Upsilon(p)=0,
\end{equation}
with $\gamma^{i}$ the $4\times 4$ Dirac matrices, which is invariant
under $e\rightarrow -e$, but not under $T\rightarrow -T$. Indeed,
just as for antimatter in momentum representation, the negative
time states could not be physically ignored, since there is nothing
to prevent a system from making a transition from a state of positive
time to a state of negative time.

Extending the above formalism to noninertial momentum-energy reference
systems on which general covariance is imposed [2], we are left with an
invariant momentum interval
\begin{equation}
dn=f_{ij}dp^{i}dp^{j},
\end{equation}
where, analogously to metric tensor $g_{ij}$ with respect to
space-time and gravitational field, relative to an arbitrarily
chosen system of momentum-energy coordinates, the functions $f_{ij}$
describe the metrical relations of the vacuum momentum-energy
continuum and also a field which is complementary to gravity.
Such a field is what we have gained from formulating
relativity in momentum-energy representation. But, what
does this new field represent?. First of all, according
to our requeriment in (ii), it must govern either
self-interaction of the wholeness of independent
space-temporal systems evolving in vacuum, rather
than space-time, or the interaction of two or more of
such systems. Moreover, if we consistently extend all
our above requirements to also embrace noninertial
reference systems, so that it is only this new field
that constitutes all of the new physics gained, then
the whole mathematics of general relativity would exactly
re-appear for noninertial momentum-energy relativity by simply
replacing $g_{ij}$ for $f_{ij}$ and $x^{i}$ for $p^{i}$ within the
realm of Riemmanian geometry, and finally we shall arrive
to obtain field equations exactly like Einstein equations,
but expressed in terms of $f_{ij}$, $p^{i}$ and the
{\it space-time} 4-tensor, rather than the usual $g_{ij}$,
$x^{i}$ and {\it momentum-energy} 4-tensor (see Eqn. (18)).
These equations can then be interpreted as requiring that
the space-temporal characteristics of a system evolving
in vacuum cause momentum-energy curvature $C$, or
alternatively, that such characteristics of the system
evolve in a certain way in response to the existence
of curvature $C$. Consider now two of such systems,
each causing a certain amount of vacuum curvature,
and evolving in a certain way through momentum-energy
in response to the $C$-curvature caused by the other.
The net result would be that each system influences the
other in such a way that their mutual vacuum momentum
separation will always tend to decrease. It is this
influence that describes the complementary field.
Moreover, if the mutual influence is very small (large), then
the vacuum momentum separation between the two systems
will become huge (tiny) and, by the uncertainty principle, their
mutual space distance very small (large). Hence, the stronger
gravity, the weaker its complementary field, and {\it vice versa}.
It follows that the new field would manifest as a repulsive interaction
in space-time. It appears then that the new interaction is
interpretable as a {\it cosmological} field which influences
every possible system evolving in vacuum momentum-energy
continuum. The ultimate evolution of a given system
will be the result of the competition between the gravitational
and cosmological interactions to which it is subject.

Of most interest is the
expression which describes the evolution of a system in a
cosmological field with potentials $f_{ij}$, i.e.
\begin{equation}
\frac{d^{2}p^{i}}{dn^{2}}+\Upsilon^{i}_{kl}u^{k}u^{l}=0,
\end{equation}
with $\Upsilon^{i}_{kl}$ the complementary affine connection in
momentum-energy
\begin{equation}
\Upsilon^{i}_{kl}=\frac{1}{2}f^{im}(\frac{\partial f_{mk}}{\partial p^{l}}
+\frac{\partial f_{ml}}{\partial p^{k}}-\frac{\partial f_{kl}}{\partial p^{m}})
\end{equation}
and the $u^{i}$ are the conventional 4-velocities in space-time which equal
their counterparts $s^{i}$ for momentum-energy.
We shall denote the momentum-energy itinerary followed by the system
as a {\it cosmodesic}.

By a suitable choice of the momentum
system we can always make all $\Upsilon^{i}_{kl}=0$ at an arbitrary
value of momentum-energy, and this would mean the elimination of the
cosmological field in the given infinitesimal element of momentum-energy.
At such an element, it would happen that the gravitational field
becomes infinite, and {\it vice versa}, at the space-time points
where the gravitational field is made to vanish the cosmological
field would diverge. This discriminated singularities of two
essentially different kinds: spacetime singularities at which
gravity becomes infinite and the cosmological field vanishes, and
momentumenergy singularities at which gravity vanishes and the
cosmological field diverges.

On the other hand, one would have as cosmological field equations
\begin{equation}
C_{ik}=\frac{4\pi K}{3}(S_{ik}-\frac{1}{2}f_{ik}S),
\end{equation}
where $C_{ik}$ is the Ricci curvature tensor as expressed in terms
of $f_{ik}$ and $p^{i}$, instead of $g_{ik}$ and $x^{i}$, respectively,
$K$ is a cosmological coupling constant having the dimension of a
coventional force, and $S$ is a {\it space-time} 4-tensor. A more detailed
study of the cosmological field and its relation with gravitation will
be considered elsewhere. Here, we shall only regard the
cosmological field when all involved velocities are small compared
with the velocity of light. In such a case, the components of $S^{k}_{i}$
are given by
\begin{equation}
S_{i}^{k}=\tau_{0}u_{i}u^{k},
\end{equation}
with $\tau_{0}$ the time momentum-density of the system, i.e. the
time that characterizes the system in a unit momentum-volume,
$\tau_{0}=\frac{T_{0}}{V_{p}}=\frac{3T_{0}}{4\pi p^{3}}$. Let us
introduce then a nonrelativistic cosmological field and denote it by
$\rho$. In general, such a field will depend on coordinates $p$ and $e$
and have the dimension of the inverse of a squared velocity. The
nonrelativistic complementary Lagrangian with dimension of a time
in the presence of a cosmological field will be given by
\begin{equation}
\tilde{L}\approx T_{0}(\frac{v^{2}}{2c^{2}}-c^{2}\rho-1)
\end{equation}
and is associated with a complementary nonrelativistic action
\begin{equation}
P\approx -T_{0}c\int de(\frac{1}{c}+c\rho-\frac{v^{2}}{2c^{3}}).
\end{equation}
It follows that in this limiting case the components of the complementary
metric tensor become
\begin{equation}
f_{00}=c^{2}(c^{2}\rho^{2}+2\rho), f_{\alpha\alpha}=1+c^{2}\rho , \alpha=1,2,3
\end{equation}

In the limit $\frac{v}{c}\rightarrow 0$, the components of 4-velocity
become $u^{\alpha}=0$, $u^{0}=u_{0}=1$, so that the only surviving
component of tensor $S^{k}_{i}$ is $S^{0}_{0}=\tau_{0}$. From (18) we
then obtain
\[C^{0}_{0}=\frac{4\pi K}{3}\tau_{0} . \]
We note that the terms containing derivatives of the affine connections
$\Upsilon^{i}_{kl}$
with respect to $\frac{e}{c}$ involve extra power $c$ and, unlike for Einstein
gravity,
are large as compared to the derivative with respect to momenta $p^{\alpha}$.
Therefore the term that dominates in the nonrelativistic limit the
Ricci curvature tensor
\begin{equation}
C^{0}_{0}=\frac{\partial\Upsilon^{\alpha}_{00}}{\partial
p^{\alpha}}-\frac{\partial\Upsilon^{\alpha}_{0\alpha}}{\partial p^{0}}
+\Upsilon^{\alpha}_{00}\Upsilon^{\beta}_{\alpha\beta}-\Upsilon^{\beta}_{0\alpha}\Upsilon^{\alpha}_{0\beta}
\end{equation}
is the negative second term on the r.h.s. It follows then that in
this limit $C^{0}_{0}\approx
-\frac{\partial\Upsilon^{\alpha}_{0\alpha}}{\partial p^{0}}$.
Now, unlike for the gravitational field [3], one should expect that the
assumption of small velocities implies the requirement that the
cosmological field be strong, for if this were not the case an
individual subsystem located in it would acquire a
high arbitrary velocity with respect to the whole system,
which is contrary
to the very notion of the cosmological field whose effect
would instead be slowing all individual independent subsystem motions
to produce an overall joint evolution pattern. Therefore, one expected
that when $v\rightarrow c$ the cosmological field would vanish, and
for small velocities $c^{2}\rho\gg 1$. Hence
\begin{equation}
\Upsilon^{\alpha}_{0\alpha}=\frac{1}{2}f^{\alpha\beta}\frac{\partial
f_{\alpha\beta}}{\partial p^{0}}
\approx -\frac{1}{2}c\frac{\dot{\rho}}{\rho},
\end{equation}
where $\dot{\rho}=\frac{d\rho}{de}$. We obtain finally
\begin{equation}
\frac{d}{de}(\frac{\dot{\rho}}{\rho})\approx\frac{8\pi K\tau_{0}}{3c^{2}}.
\end{equation}
If the system is characterized by a single time $T_{0}$, then we obtain
an asymptotic expression for the cosmological field
\begin{equation}
\rho\approx\rho_{0}\exp(\frac{KT_{0}c^{2}}{Pv^{2}}),
\end{equation}
where we have used $e^{2}=\frac{P^{2}c^{4}}{v^{2}}$, and $\rho_{0}$ is an
integration constant. Potential (26) would, as expected, correspond
to a repulsive field. Note that the larger the "momentum separation",
the smaller the value of the
potential, and the larger the absolute value of the involved
gravitational potential.

Let us consider now the case of a constant cosmological field, meaning
by that a cosmological field which does not depend on vacuum energy
coordinate $p^{0}$: $f_{ik}\neq f_{ik}(p^{0})$. In this case the
coordinate $p^{0}$ should be a true {\it vacuum energy} relating to
the complementary proper energy $e$ by $e=cf_{00}^{\frac{1}{2}}p^{0}$.
In the case of a strong cosmological field this relation would reduce
to
\begin{equation}
e\approx cp^{0}(1+c^{2}\rho)
\end{equation}
for which $c^{2}\rho\gg 1$ and the first of expressions (22) have
been used.

Let us look then at
the momentum-energy propagation in a large constant cosmological
field of a light ray with an {\it energy frequency} $\Omega$ [4] which
would generally be given by the derivative with respect to
energy coordinate of the phase-eikonal in momentum-energy
continuum $\varphi=-r_{i}p^{i}+\beta$ (where $\beta$ is an arbitrary
constant)
for a momentum-energy "plane
wave" $g=g_{0}e^{i\varphi}$ which corresponded to just one
space-time point. As expressed in terms of the vacuum energy $p^{0}$,
the energy frequency becomes $c\Omega_{0}=-\frac{\partial\varphi}{\partial
p^{0}}$,
and if we express it in terms of the proper energy $e$, then we
have
\begin{equation}
c\Omega=-c\frac{\partial\varphi}{\partial
e}=-f_{00}^{-\frac{1}{2}}\frac{\partial\varphi}{\partial p^{0}}
\approx\frac{\Omega_{0}(1-\frac{1}{c^{2}\rho})}{c\rho}.
\end{equation}
Thus, if a ray of light is emitted at a point where the cosmological
potential is $\rho_{1}$ and the energy frequency is $\Omega$, then
upon arriving at a point where the potential is $\rho_{2}$ it will
have an approximate energy frequency $\frac{\Omega\rho_{1}}{\rho_{2}}$.
For an observer at the arriving point the energy frequency would then
be shifted by an amount
\begin{equation}
\bigtriangleup\Omega\approx\Omega(\frac{\rho_{1}}{\rho_{2}}-1),
\end{equation}
which corresponded to a cosmological proper-energy shift given by
\begin{equation}
\bigtriangleup E\approx (\frac{\rho_{2}}{\rho_{1}}-1)E,
\end{equation}
where $E$ is the proper energy at the emission point where the
cosmological potential is $\rho_{1}$.

In order to analyse the possible observational meaning of (30), let
us consider the light coming to our galaxy from another luminous
galaxy. It appears that the light emitted from inner regions of
the latter galaxy should be produced in a physical
environment where the local gravitational velocities would generally
be comparable to those occurring in our own; therefore, for such a
light, both the gravitational and cosmological fields would be
expected to be of the same intensity for the two galaxies, and
$\bigtriangleup E\approx 0$. However, as the light source separates
from the core and enters the outer regions of the emitting galaxy,
its gravitational atmosphere became weaker and hence its related
velocities would be smaller. Thus, the cosmological field at the
outskirts of the emitting galaxy should be expected to be stronger
than at the receiving point on our galaxy. One then would generally
expect $\bigtriangleup E < 0$ for light coming from galactic halos.

An observer in our galaxy who were unaware of (30)
would therefore interpret the amount of
luminous matter in the emitting galaxy to become smaller than it
actually is as the source is shifted from galactic centers first towards
the halos, and then deep into the intergalactic media where the
cosmological field became strongest. It appears that, rather than
interpreting this indeed observed effect as caused by the existence
of more or less exotic dark matter [5], it should be attributed to the
cosmological shift (30) which originates from the noninvariance of
the proper energy under propagation in momentum-energy continuum.

On the other hand, one would not exclude the observational emergence
of just the opposite effect; i.e. the effect arising from physical
situations where the cosmological field at the source is weaker than
in our galaxy. Such situations would in principle appear when the
emitting regions possessed very strong gravitational fields and
correspond to positive cosmological shifts, $\bigtriangleup E > 0$.
An observer in our galaxy would then attribute an anormally high
luminosity to the source, typically perhaps an active galactic nucleus
or a quasar. Of course, all the above discussion is rather qualitative
as the approximate expression for the cosmological energy shift given
by (30) only is strictly valid for cosmological fields sufficiently
strong as for $c^{2}\rho$ to still be much greater than unity.

Extrapolating to the point where all velocities vanish, a true
momentum-energy singularity would be reached where
$\rho\rightarrow\infty$ and
$\bigtriangleup E=-E$. So, no luminous matter could be detected
from that point by any observers at any regions. If we assumed
this to be big bang singularity, then inmediately after it the
huge value of the repulsive potential $\rho$ given by (26) would
drive the universe through a very short-lived phase of exponential
expansion which rapidly damped off to smoothly enter a regime
where the cosmological potential is small enough to allow dominance
of Einstein gravitational field, and hence of an evolution that
proceeds in accordance with conventional Friedmann expansion. All
memory of the cosmological field left on the universe as a whole
would then be slowly erased as expansion goes on, and correspond
to the usual cosmological term $\Lambda$.

\vspace{1cm}

{\bf Acknowledgements}. The author is grateful to {\it CAICYT} for
support under Research Project N§ PB91-0052. I also thank Carmen
Siguenza for her useful comments, constructive criticisms and
constant encouragment, and Jos\'e Mar\'{\i}a G\'omez Soto for his
interest and deep understanding.

\pagebreak

\noindent\section*{References}
\begin{description}
\item [1] P.F. Gonz\'alez-D\'{\i}az, Phys. Lett. B317 (1993) 36.
\item [2] A. Einstein,{\it The Meaning of Relativity} (Chapman and Hall,
Bristol, 1967); in: {\it The Principle of Relativity} (Dover, New York,
1952).
\item [3] L.D. Landau and E.M. Lifshitz, {\it The Classical Theory of
Fields} (Pergamon Press, Oxford, 1975).
\item [4] What plays the role of a space-time wave in momentum-energy
should be a phenomenon repeating itself along the energy
coordinate, with energy frequency $\Omega$. Light signals
propagated in momentum-energy at the maximum rate $\frac{1}{c}$,
and manifested either as waves with frequency $\Omega$, or
as discrete {\it time quanta} where time concentrated so
that $T=h\Omega$. These definitions are necessary for a
consistent quantization of systems in momentum-energy,
since only they may allow that a periodic phenomenon
occurring in momentum-energy satisfies the counterpart
for complementary relativity of the de Broglie's
theorem of phase harmony [L. de Broglie,
{\it Recherches sur la Theorie des Quanta} (Masson
and Cie., Paris, 1963)]. Clearly, by using these
definitions, frequency $\Omega_{1}=\frac{T_{0}\gamma}{h}$
(from (4)) relative to a fixed observer will always keep
itseft in phase ($\varphi=\frac{T_{0}\gamma e}{h}$) with
a wave of energy frequency $\Omega=\frac{T_{0}}{h\gamma}$
(from (7)), for such an observer. The latter wave
propagated at a rate $\frac{1}{v}>\frac{1}{c}$ and
carried only phase. It can be also shown that the
group rate for such phase waves is $\frac{v}{c^{2}}$,
i.e. just the propagation rate of the system in
momentum-energy.
\item [5] P.J.E. Peebles, {\it Principles of Physical Cosmology}
(Princeton University Press, Princeton, 1993).

\end{description}

\end{document}